\documentclass[a4paper,12pt]{article}

\usepackage[utf8]{inputenc}
\usepackage{epsfig}
\usepackage{graphicx}
\usepackage{amsmath}
\usepackage{amssymb}
\usepackage{cmap}
\usepackage{bm}
\usepackage{hyperref}
\usepackage{array}
\usepackage{epsfig}
\usepackage{array}
\usepackage{color}
\usepackage{latexsym}
\usepackage{xypic}
\usepackage{amscd}
\usepackage{lipsum}

\usepackage{cite} 

\usepackage[left=0.85in, right=0.85in, top=0.85in, bottom=0.9in]{geometry}

\newcommand{\version}{\sf arXiv:~v4\ ---\ \today} 

\title{\huge Doubly charmed baryon mass and wave function through a random walks method \thanks{\version}}

\author{B.O.~Kerbikov$^{a,b,c,}$\thanks{borisk@itep.ru}\, \\
\\
$^a$ \small{\em Alikhanov Institute for Theoretical and Experimental Physics,}\\
\small{\em B. Cheremushkinskya 25, 117218 Moscow, Russia}\vspace{0.25cm}\\
$^b$ \small{\em Lebedev Physical Institute,}\\
\small{\em Leninsky Prosp. 53, 117924 Moscow, Russia}\vspace{0.25cm}\\ 
$^c$ \small{\em Moscow Institute of Physics and Technology,}\\
\small{\em Institutskiy per. 9, 141700 Dolgoprudny, Moscow Region, Russia}
}\bigskip
\date{\today}

\newcommand{\be}{\begin{equation}}
\newcommand{\ee}{\end{equation}}

\def\fun#1#2{\lower3.6pt\vbox{\baselineskip0pt\lineskip.9pt
\ialign{$\mathsurround=0pt#1\hfil ##\hfil$\crcr#2\crcr\sim\crcr}}}

\newcommand{{\SD}}{\rm SD}

\newcommand{{\Mc}}{\mathcal{M}}

\newcommand{\PRLsep}{\noindent\makebox[\linewidth]{\resizebox{0.4\linewidth}{1.2pt}{$\bullet$}}\bigskip}

\begin{document}
\maketitle

 
\begin{abstract}
{

\noindent The mass and the wave function of doubly charmed $\Xi^{++}_{cc}$ \textit{(ccu)} baryon are evaluated using Green Function Monte Carlo method to solve the three-body problem with Cornell potential. The mass of $\Xi^{++}_{cc}$ with spin $1/2$ is in a good agreement with the LHCb value. Simulation of the wave function by random walks resulted in a configuration of the quark-diquark type. The radius of $\Xi^{++}_{cc}$ is much larger than the size needed for a large isospin splitting. The prediction for the $\Omega_{cc}$ mass is presented.

}
\end{abstract}\bigskip

The first detection of a baryon $\Xi^{++}_{cc}$ containing two charm quarks (i.e., having the structure {\it ccu}) has been made at CERN by LHCb collaboration last summer \cite{01}. Its mass was measured to be $3621$ MeV. The existence of such a baryon may be considered is an inevitable consequence of the existence of the $c$-quark itself. Calculations of the $\Xi^{++}_{cc}$-baryon mass have started almost thirty years ago \cite{02,03,04}. Probably the first review paper on this subject dates back to 2002 \cite{05}. A reliable prediction on the double charmed baryon mass and properties turned out to be a difficult talk -- see a list of references in \cite{01}. This is not surprising since the theoretical description of the QCD ``hydrogen atom'' -- charmonium, to a great extent relies on the phenomenological input rather than on the first principles of QCD. Connection between different approaches to the double charmed baryons, e.g., the quark-diquark model, the potential model, and the QCD sum rules, their interreletion and relation to the fundamental QCD Lagrangian are obscure. The aim of the present note is to investigate the $\Xi^{++}_{cc}$ mass and the structure of the wave function taking our works \cite{03,04} as a starting point for a discussion. In \cite{03,04} the spin-averaged masses and the wave functions of multiquark systems made up to 12 quarks were calculated in potential model through Green Function Monte Carlo, or random walks method -- see below. 

Despite the fact that QCD gives no sound arguments in favor of the effective potential with only two-body forces, the constituent quark model has given results which are in surprisingly good agreement with experimental hadron spectroscopy. Attempts to apply this approach to multiquark systems encounter serious difficulties even if one ignores problems like complicated QCD structure in the infrared region. With growing number of particles, especially with non-equal masses, and for the complicated form of the potential, the traditional methods -- variational, integral equations, hyper-spherical functions, face difficulties. The accuracy, in particular that of the many-body wave function determination, becomes uncontrollable and the computation time increases catastrophically. This is true even for the three quark system. The convergence for the wave function is much slower than for the binding energy both in the harmonic oscillation expansion and in the hyperspherical formalism \cite{02}. In \cite{03,04} the spectrum and the wave functions of multiquark systems were investigated using the Green Function Monte Carlo method (GFMC). The GFMC is based on the idea attributed to Fermi that the imaginary time Schrodinger equation is equivalent to the diffusion equation with branching (sources and sinks).  {The GFMC allows to calculate easily and with high accuracy the spectrum and, what is most important, the wave function of a multiquark system in various models.} Originally GFMC has been used to calculate the ground-state properties of a variety of systems in statistical and atomic physics \cite{06}. At present it is the most powerful method to solve the many-body problem \cite{07}. Similar methods have been used in Quantum Field Theory under the names Projector Monte Carlo \cite{08} and Guided Random Walks \cite{09}.
The description of the GFMC is beyond the scope of this paper. We only stress that the method does not require to solve differential or integral equations for the wave function. There is no need even to write down such equations. In GFMC the exact many-body Schrodinger equation is represented by a random walk in the many-dimensional space in such a way that physical averages are exactly calculated given sufficient computational resources. {At this point we emphasize the need to disentangle the accuracy of computations from possibly large uncertainty related to the use of a particular model.} When we tried to apply GFMC to multiquark system, i.e., to a system of fermions, we encounter a difficulty. For the fermionic system the kernel of the Green function may take a negative value at a certain step of the sampling procedure. A recipe to circumvent this difficulty was proposed in \cite{03}. We note in passing that similar problem does not allow to perform lattice Monte Carlo investigation of quark matter properties at finite density.

Now we come to the description of the interquark interaction chosen for the calculation of the ground state mass of \textit{(ccu)} system and evaluation of the wave function. For a system of $N$ quarks the interaction between quarks is taken in the form 
\be
V_{ij}(r_{ij})=\lambda_i\lambda_j V_{8}(r_{ij}),
\label{eq:01}
\ee

{
\noindent with $\lambda_i$ being the Gell-Mann matrices. Solving the quark problem with $N \geq 4$ with the interaction (\ref{eq:01}) one encounters a problem that the coupling of colors of the constituents into a color singlet is not unique -- see \cite{04} for a discussion. For $V_{8}(r)$ we have taken the well-known Cornell potential \cite{10}
\be V_{8}(r)=-\frac{3}{16}\left(-\frac{\varkappa}{r}+\frac{r}{a^2}+C_f  \right), \label{eq:02} \ee
where $\varkappa=0.52$ and $a=2.34$ GeV$^{-1}$ which corresponds to the string tension $\sigma = a^{-2} = 0.18$ GeV$^2$. For the baryon $\left<\lambda_i\lambda_j\right>=-8/3$. We would like to present two more arguments in addition to the well known ones \cite{10} in support of the Cornell potential. Slightly varying its parameters one may obtain a good fit of the lattice simulation of the quark-antiquark static potential \cite{11}. Cornell like behavior arises also from AdS/QFT correspondence \cite{12}.}


In Ref. \cite{13} an excellent fit of both meson and baryon sectors has been obtained under the assumption that the constant $C_f$ is weakly flavor dependent. Following \cite{13} we have chosen the input parameters for \textit{(ccu)} baryon (all values in GeV units)
\be
m_u=0.33,\,\,\,m_c=1.84,\,\,\,C_{uc}=-0.92.
\label{eq:03}
\ee
The value of $m_c$ might look too high but it corresponds to the Classical Cornell set of parameters \cite{10}.

Next one has to evaluate the contribution from spin-spin splitting \cite{14}
\be
V_{s_i s_j} = \frac{16\pi}{9}\,\alpha_S\,\frac{\bm{s}_i\bm{s}_j}{m_i m_j}\,\delta^{(3)}(\bm{r}_{ij}).
\label{eq:04}
\ee
We have calculated the ground-state expectation values $\delta_{ij} = \left<\delta(\bm{r}_{ij})\right>$, where $\left<\, \ldots \,\right>$ means the average over the ground-state wave function. To obtain $\delta_{ij}$ we made a smearing over the small sphere around the origin and then averaged over a sequence of such spheres. For \textit{(ccu)} baryon we obtain the following results for $10^3\,\delta_{ij}$ (GeV$^3$)
\be
\delta_{cc}=45.25 \pm 2.90,\quad  \delta_{cu}=11.28 \pm 1.94.
\label{eq:05}
\ee
The GFMC method allows to obtain the wave function and its arbitrary moments with the accuracy restricted only by the computational resources. In particular, there is no need to introduce the quark-diquark structure as a forced anzatz. The system will form such configuration by itself if it corresponds to the physical picture. This turned out to be the case for \textit{(ccu)} baryon. Indeed, the ground-state expectation values of $\left< r_{ij}^{2} \right>^{1/2}$ in GeV$^{-1}$ are
\be
\left< r_{cc}^{2} \right>^{1/2} = 2.322\pm 0.024,\quad
\left< r_{cu}^{2} \right>^{1/2} = 3.407 \pm 0.035.
\label{eq:06}
\ee
If we identify $\left< r_{cc}^{2} \right>^{1/2} \simeq 0.46$ fm with the size of a diquark, then it is more compact one than a diquark with $r_d=0.6$ fm introduced ad hoc in \cite{15}. {It is instructive to look at the interquark distances (\ref{eq:06}) from an angle of the $\Xi^{++}_{cc} - \Xi^{+}_{cc}$ isospin splitting. This is a long standing puzzle. More than a decade ago SELEX Collaboration reported \cite{16} the observation of the $\Xi^{+}_{cc}$ \textit{ccd} baryon with a mass $3519$ MeV. However, this result was not confirmed by other experiments (see \cite{01} for references). The isospin splitting of about $100$ MeV between \textit{ccu} and \textit{ccd} states and its sign are hardly possible to explain. The $d$-quark is heavier than the $u$-quark and to overcome the ``wrong'' sign of the splitting by about $100$ MeV the electromagnetic mass difference should be very large. This is turn requires the $\Xi^{++}_{cc}$ baryon to be very compact \cite{17}. To obtain $9$ MeV splitting the radius should satisfy $\sqrt{\left< r^2 \right>} < 0.26$ fm \cite{17}. From (\ref{eq:06}) we see that the diquark size is $0.46$ fm. If we identify the baryon radius with three quarks hyperradius $\rho = \sqrt{\bm{\eta}^2 + \bm{\xi}^2}$ with $\bm{\eta}$ and $\bm{\xi}$ being the Jacobi coordinates \cite{02}, the result is $\rho=(0.53-0.57)$ fm depending on the value of the Delves angle $\tan \varphi = {\xi}/{\rho}$. Therefore our result on the $\Xi^{++}_{cc}$ wave function strongly contradicts the abnormal value and the sign of the conjectural isospin splitting.} 

{
Our result for the center of gravity (spin-averaged)  mass $\Xi^{++}_{cc}$ baryon is \be m\left[ \Xi^{++}_{cc} \right] = 3632.8 \pm 2.4 \text{ MeV}.\label{eq:07} \ee
The error characterizes the accuracy of the GFMC calculations with the Cornell potential parameters specified above. We did not vary these parameters since they were fitted to a great body of observables. Next we take into account the hyperfine interaction (\ref{eq:04}) which induces the splitting between the lowest state with $S=1/2$ and its $S=3/2$ partner.

Taking into account that statistics requires the \textit{cc} pair to be in a spin $1$ state, we can write the following equation for the hyperfine energy shift 
\be
\Delta E_{hf} = \frac{4\pi}{9}\,\alpha_S\,\left\{\frac{\delta_{cc}}{m_{c}^2} + \frac{2\delta_{cu}}{m_{c}m_{u}}\left[ S(S+1) - \frac{11}{4}\right]\right\}.
\label{eq:08}
\ee
We used the Cornell value of $\alpha_S=\frac34\,\varkappa=0.39.$ With this value of $\alpha_S$ baryon magnetic moments were successfully described \cite{18}. Equation (\ref{eq:08}) yields 
\be \Delta E_{hf}\left(S=1/2\right)=-32.2\text{ MeV}, \qquad \Delta E_{hf}\left(S=3/2\right)=26.9\text{ MeV}. \label{eq:09} \ee and \be 
m\left[ \Xi^{1/2\,++}_{cc} \right] = 3601 \text{ MeV},\qquad m\left[ \Xi^{3/2\,++}_{cc} \right] = 3660 \text{ MeV}.\label{eq:10} \ee
In (\ref{eq:10}) the uncertainty of GFMC calculations which are about $(2-3)$ MeV are not presented since as repeatedly stated above the model-dependent theoretical uncertainty may be much larger as one can see from theoretical predictions presented in the reference list in \cite{01}.   

Another doubly charmed baryon which may be observed soon is $\Omega_{cc}$ with quark content {(ccs)}. Our result for its c.o.g. is \be m\left[ \Omega_{cc} \right] = 3760.7 \pm 2.4 \text{ MeV}.\label{eq:11} \ee
}


\PRLsep

The author is grateful to Yu.S.~Kalashnikova for enlightening discussions, to M.~Karliner, V.~Novikov, J.-M.~Richard, and M.I.~Vysotsky for questions and remarks. The interest to the work fron V.Yu.~Egorychev is gratefully acknowledged. The work was supported by the grant from the Russian Science Foundation project number \#16-12-10414. 





\begin{thebibliography}{99}

\bibitem{01} 
R.~Aaij, B.~Adeva, M.~Adinolfi,  et al. \textit{(LHCb Collaboration)}, Phys. Rev. Lett. \textbf{119} (2017) 112001 [arXiv:1707.01621 [hep-ex]].

\bibitem{02}
S.~Fleck and J.-M.~Richard, Progr. Theor. Phys. \textbf{82} (1989) 760.

\bibitem{03} 
B.O.~Kerbikov, M.I.~Polikarpov, L.V.~Shevchenko, and A.B.~Zamolodchikov, Yad. Fiz. \textbf{46} (1987) 886.

\bibitem{04} 
B.O.~Kerbikov, M.I.~Polikarpov, and L.V.~Shevchenko, Nucl. Phys. \textbf{B 331} (1990) 19.

\bibitem{05}
V.V.~Kiselev and A.K.~Likhoded, Phys. Usp. \textbf{45} (2002) 455 [arXiv:hep-ph/0103169]. 

\bibitem{06} 
M.H.~Kalos, Phys. Rev. \textbf{128} (1962) 1791;

D.~Ceperly and B.~Alder, Science \textbf{231} (1986) 555;

D.~Ceperly and B.~Alder, J. Chem. Phys. \textbf{81} (1984) 5833.

\bibitem{07} 
J.~Carlson, S.~Gandolfi, F.~Pederiva, et al., Rev. Mod. Phys. \textbf{87} (2015) 1067. 

\bibitem{08} 
R.~Blankenbecler and R.L.~Sugar, Phys. Rev. \textbf{D 27} (1983) 1304. 

\bibitem{09} 
S.A.~Chin, J.W.~Negele, and S.E.~Koonin, Ann. Phys. \textbf{157} (1984) 140.

\bibitem{10}
E.~Eichten, K.~Gottfried, T.~Kinoshita, et al., Phys. Rev. \textbf{D 17} (1978) 3090; Phys. Rev. \textbf{D 21} (1980) 313 \textit{(Erratum)};

E.~Eichten, C.~Quigg, Phys. Rev. \textbf{D 52} (1995) 1726 [arXiv:hep-ph/9503356];

E.~Eichten, S.~Godfrey, H.~Mahlke, and J.L~Rosner, Rev. Mod. Phys. \textbf{80} (2008) 1161 [arXiv:hep-ph/0701208].

\bibitem{11} 
G.S.~Bali, Phys. Rep. \textbf{343} (2001) 1 [arXiv:hep-ph/0001312]. 

\bibitem{12}
C.D.~White, Phys. Lett. \textbf{B 652} (2007) 79. 

\bibitem{13} 
A.M.~Badalyan and D.I.~Kitaroage, Yad. Fiz. \textbf{47} (1988) 1343.

\bibitem{14} 
A.~De~Rujula, H.Georgi, and S.L.~Glashow, Phys. Rev. \textbf{D 12} (1975) 147.

\bibitem{15} 
V.V.~Kiselev, A.V.~Berezhnoy and A.K.~Likhoded, arXiv:1706.09181 [hep-ph].

\bibitem{16} 
M.~Mattson, G.~Alkhazov, A.G.~Atamantchouk, et al., Phys. Rev. Lett. \textbf{89} (2002) 112001 [arXiv:hep-ex/0208014]; 

A.~Ocherashvili, M.A.~Moinester, J.~Russ, et al., Phys. Lett. \textbf{B 628} (2005) 18 [arXiv:hep-ex/0406033].

\bibitem{17} 
S.J.~Brodsky, Feng-Kun~Guo, C.~Hanhart, Ulf-G.~Mei\ss ner, Phys. Lett. \textbf{B 698} (2011) 251 [arXiv:1101.1983 [hep-ph]].

\bibitem{18} 
B.P.~Kerbikov and Yu.A.~Simonov, Phys. Rev. \textbf{D 62} (2000) 093016 [arXiv:hep-ph/0001243].

\end{thebibliography}
\end{document}